# Observation of Andreev bound states in bicrystal grain-boundary Josephson junctions of the electron doped superconductor $La_{2-x}Ce_xCuO_{4-y}$

B. Chesca [1*], M. Seifried [2], T. Dahm [2], N. Schopohl [2], D. Koelle[1], R. Kleiner [1], A. Tsukada[3]

1: Physikalisches Institut-Experimentalphysik II, Universität Tübingen,
Auf der Morgenstelle 14, D-72076 Tübingen, Germany
2: Institut für Theoretische Physik, Universität Tübingen,
Auf der Morgenstelle 14, D-72076 Tübingen, Germany
3: NTT Basic Research Laboratories, 3-1 Morinosato Wakamiya,
Atsugi-shi, Kanagawa 243, Japan

We observe a zero-bias conductance peak (ZBCP) in the *ab*-plane quasiparticle tunneling spectra of thin film grain-boundary Josephson junctions made of the electron doped cuprate superconductor $La_{2-x}Ce_xCuO_{4-y}$. An applied magnetic field reduces the spectral weight around zero energy and shifts it non-linearly to higher energies consistent with a Doppler shift of the Andreev bound states (ABS) energy. For all magnetic fields the ZBCP appears simultaneously with the onset of superconductivity. These observations strongly suggest that the ZBCP results from the formation of ABS at the junction interfaces, and, consequently, that there is a sign change in the symmetry of the superconducting order parameter of this compound consistent with a d-wave symmetry.

Crucial to the successful development of a microscopic theoretical model for superconductivity in the high transition temperature cuprates (HTS) is the knowledge of the symmetry of the order parameter describing the pairing of electrons in the superconducting state. Whereas for hole doped cuprate superconductors the $d_{x^2-y^2}$ (*d*) -wave symmetry has been established, for electron doped HTS $R_{2-x}Ce_xCuO_4$ *(RCCO* with *R= La* (L), *Nd* (N), *Pr* (P), *Sm, Eu*) where carriers are predominantly electrons, the issue remains controversial [1-3]. In particular the formation of zero-energy Andreev bound states (ABS) at the junction interface [2-4], that supports *d*-wave symmetry, has been controversial as far as electron doped HTS are concerned. ABS at the Fermi energy arise from constructive interference between the electronlike and holelike quasiparticles incident and reflecting at the junction interface, which experience different signs of the order parameter. ABS [2-4] lead to a zero–bias conductance peak (ZBCP) in the quasiparticle tunneling spectra. For a d-wave superconductor a ZBCP due to ABS is expected for quasiparticle injection into the *ab* plane for all surfaces except (100) and (010). By contrast for an *s*-wave superconductor no ABS are formed. Hence, when identified as arising from ABS, the observation of a ZBCP is a clear signature for a predominant *d*-wave symmetry of the order parameter. Regarding electron-doped HTS, a ZBCP has been observed in PCCO thin film normal metal-insulator-superconductor (NIS) junction tunneling spectra [5] and in the spectra of a NCCO single crystal-normal metal junction [6] (although in this case the appearance of a ZBCP has *not* been attributed to d-wave symmetry). However, no ZBCP was present in similar measurements involving NCCO single crystals [7] or thin film NCCO NIS junctions [8]. In contrast to NIS junctions, grain boundary Josephson junctions (GBJs) provide the unique opportunity to obtain information on the symmetry of the superconducting order parameter *simultaneously* from Cooper pair Josephson tunneling and Andreev reflection of quasiparticles. Indeed, the π-phase shift in the Cooper pair tunneling spontaneously induced in a tri- or tetra-crystal geometry containing GBJs has been attributed to a *d*-wave symmetry in both hole doped and optimally electron doped cuprates. This leads to striking anomalies: spontaneous appearance of half integer magnetic flux quanta [1] or circulating currents oscillating at GHz frequencies [10], or to a magnetic field induced increase of the Josephson critical current [9,10,11]. So far, all phase sensitive tests based on the dc Josephson effect performed with GBJs of electron doped superconductors, namely with NCCO [1] and LCCO [9], supported d-wave symmetry. On the other hand, ZBCP's have been only observed for hole doped GBJs [12-14], while all previous attempts made on electron doped NCCO, PCCO and LCCO GBJs have failed so far [13-16] challenging the correctness of a presumable *d*-wave symmetry. Why Josephson tunneling on one hand and quasiparticle tunneling on the other, both performed on GBJs give contradictory results as far as electron doped cuprates are concerned? Here we present ZBCP measurements of LCCO GBJs that may reconcile these previous contradictory results.

---

* Corresponding author: boris.chesca@uni-tuebingen.de

For our experiments, 1 μm thick c-axis oriented LCCO thin films were epitaxially grown on $SrTiO_3$ substrates by molecular-beam epitaxy, as reported elsewhere [17]. The films were near optimal doping with $x = 0.105$ and a critical temperature of about $T_c = 29$ K. The $SrTiO_3$ (STO) substrates contain $30^0$ [001] tilt symmetric grain boundaries. Subsequently one film was patterned by standard photolithography and Ar ion milling to form GBJs of widths $w$ between 200-1000 μm. We made four-point measurements to obtain the current-voltage characteristics (IVCs) and numerically differentiate these to obtain the differential conductance (G = dI/dV). We current-biased the samples and measured the voltage with a resolution better than about 0.2 μV, a level settled by the environmental noise. We measured 5 GBJs patterned on this film (cf. Table 1) and in all cases a well defined ZBCP was present. In this work we selected one representative example (GBJ #1) showing the strongest ZBCP and present a comprehensive set of measurements; current-voltage characteristics in zero applied field, magnetic field dependence of the Josephson critical current, temperature and field dependence of the ZBCP, ZBCP dependence on the field orientation, and a detailed analysis of the integrated spectra measured, showing conservation of states. All figures are for this particular junction except Fig.2c which shows conductance spectra at different magnetic fields of another representative GBJ, named GBJ #2.

**TABLE 1.** Properties of the LCCO $30^0$ misoriented grain boundary junctions (GBJs) measured at 4.2 K and zero applied magnetic field: junction critical current $I_c$, return current $I_r$, junction critical current density $j_c$, junction width $w$.

| GBJ | $I_c$ (μA) | $I_r$ (μA) | $j_c$(A/cm$^2$) | $w$ (μm) |
|---|---|---|---|---|
| #1 | 23 | 15 | 11.5 | 200 |
| #2 | 18 | 14 | 9 | 200 |
| #3 | 22 | 13 | 8 | 275 |
| #4 | 32 | 6.5 | 6.4 | 500 |
| #5 | 63 | 18.5 | 6.3 | 1000 |

All GBJs we measured at temperature $T = 4.2$ K and at small applied magnetic fields (up to the mT range) have hysteretic IVCs that are well described by the resistively and capacitively shunted-junction (RCSJ) model (a representative IVC is shown in Fig.1). This behavior agrees well with many other previous reports on hole- [18] and electron-doped GBJs [19]. In the example shown in Fig.1 (data for other GBJs are given in Table 1) the $w = 200$ μm wide and 1 μm thick junction has a critical current $I_c = 23$ μA. That corresponds to a junction critical current density $j_c = 11.5$ A/cm$^2$ and a Josephson penetration depth $\lambda_J = (\Phi_0/(2\pi\mu_0(2\lambda+t)j_c))^{1/2}$ of about 65 μm. Here $t$ is the physical barrier thickness, $\lambda$ is the London penetration depth ($\lambda = 250$ nm [20] was taken to calculate $\lambda_J$), $\Phi_0$ is the magnetic flux quantum and $\mu_0$ is the vacuum permeability. The junction is therefore in the short junction limit $w < 4\lambda_J$ [1,18]. $I_c$ as a function of a small applied field $\mu_0H$ (in the μT range) parallel to the *c* axis, i.e., perpendicular to the planar junction geometry (shown in the left hand side inset of Fig.1), is shown in the right hand side inset of Fig.1. This characteristic has a shape that qualitatively resembles a Fraunhofer pattern (dotted line). The 90% modulation of $I_c$ proves a good homogeneity of $j_c$ along the junction on a scale above 1 μm. There are two main discrepancies from an ideal Fraunhofer pattern; the first is that for small fields (μT range) Josephson current does not reach zero. Such a behavior is well understood [9] in terms of small structural fluctuations present along the GB due to its nano-faceted character [18]. The second discrepancy is that the data are shifted along the magnetic field axis by a small background field [9]. It should be pointed out that at fields in the mT range or higher there was no trace of a Josephson supercurrent left on the IVC.

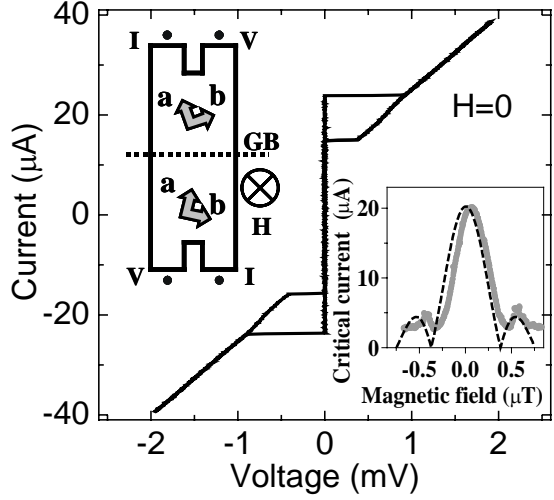

**FIGURE 1.** Current-voltage characteristic of 30⁰ GBJ #1 measured at 4.2 K in zero applied field: H=0. The left hand side inset shows the junction geometry; *a* and *b* are the unit cell vectors. The dashed line indicates the location of the grain boundary (GB). The right hand side inset shows the junction critical current as a function of the applied magnetic field (H ll *c* axis). With dotted line the theoretical Fraunhofer pattern is also shown.

Figures 2a and 2b show typical families of G(V) spectra of GBJ #1 measured for different large magnetic fields (Tesla range) applied either parallel to the *c*-axis (Fig. 2a) or parallel to the *ab* plane (and perpendicular to the grain boundary; Fig. 2b) at a temperature $T = 4.2$ K. A clear ZBCP is visible accompanied by gaplike coherence peaks at about ±9 mV. As H increases, for H ll *ab* both the width and height of the ZBCP gets suppressed faster as compared to the case H ll *c*. In contrast to the other samples measured (one such example is GBJ#2 - see Fig.2c) for the GBJ shown in Figs. 2a and 2b there are some additional structures on the ZBCP that are gradually suppressed by an increasing H or T. Thus, at 4.2 K (see Figs.2a and 2b) the structures vanish at 1.5 T when H ll *c* and at 0.8 T for H ll *ab*. Then, at 10 K (see Figs.3a and 3b) a smooth shoulder on the ZBCP is only left that vanishes as well above 1 T for H ll *c* and above 0.7 T for H ll *ab*. These fine structures which we believe are related to ABS energy shifts in magnetic fields, are not at the focus of this paper. For comparison in Fig.2c we present conductance spectra of GBJ#2 at 4.2 K and 5 different values of H ll *c* axis. To focus on the ZBCP the low voltage part of the reduced conductance $G(V)-G_{SG}(V=0)$ is plotted. The dotted line in the inset is a fit of the subgap background conductance $G_{SG}$. The extrapolation of the dotted line to V=0 yields $G_{SG}(V=0)$.

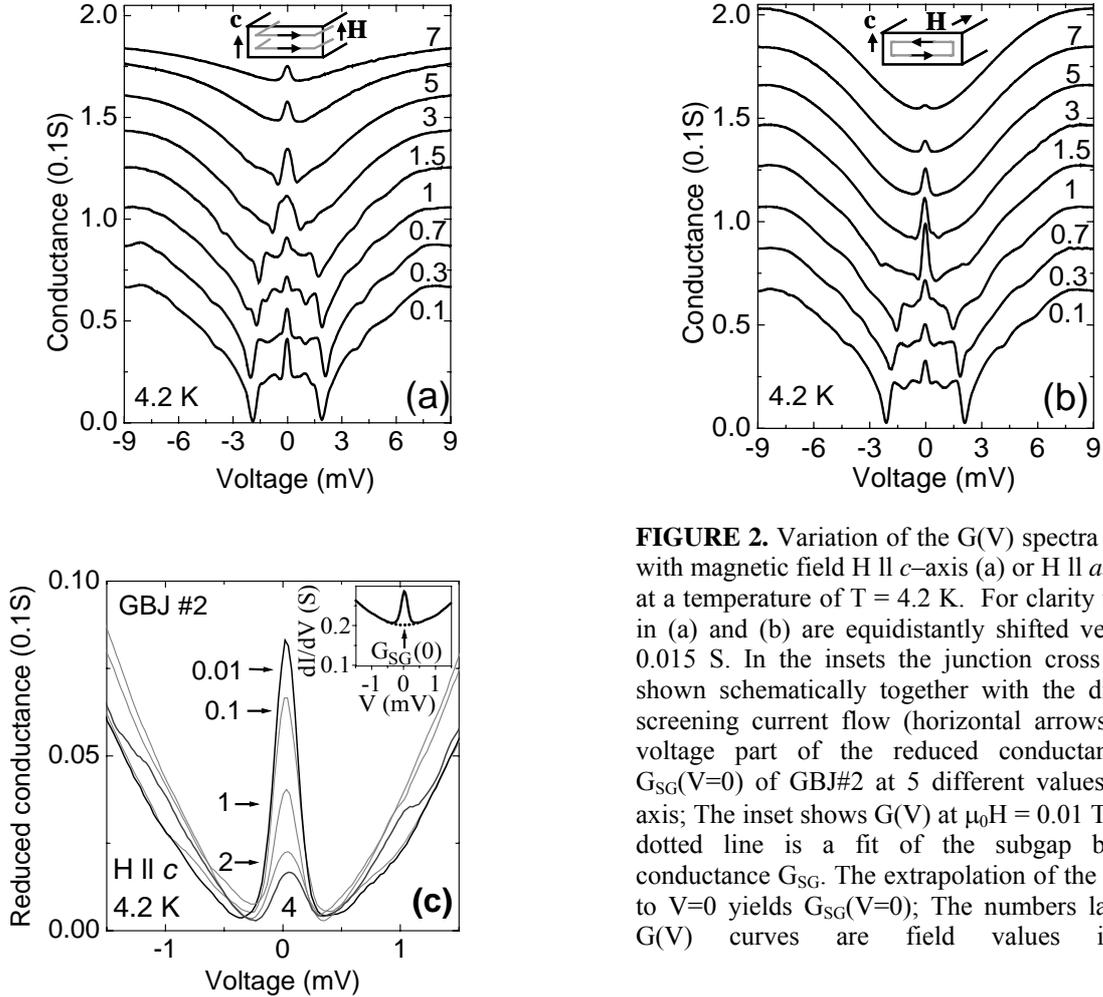

**FIGURE 2.** Variation of the G(V) spectra of GBJ #1 with magnetic field H ll *c*–axis (a) or H ll *ab* plane (b) at a temperature of T = 4.2 K. For clarity the spectra in (a) and (b) are equidistantly shifted vertically by 0.015 S. In the insets the junction cross section is shown schematically together with the direction of screening current flow (horizontal arrows). c) Low voltage part of the reduced conductance G(V)-$G_{SG}$(V=0) of GBJ#2 at 5 different values of H ll *c* axis; The inset shows G(V) at $\mu_0$H = 0.01 T. Here, the dotted line is a fit of the subgap background conductance $G_{SG}$. The extrapolation of the dotted line to V=0 yields $G_{SG}$(V=0); The numbers labeling the G(V) curves are field values in Tesla.

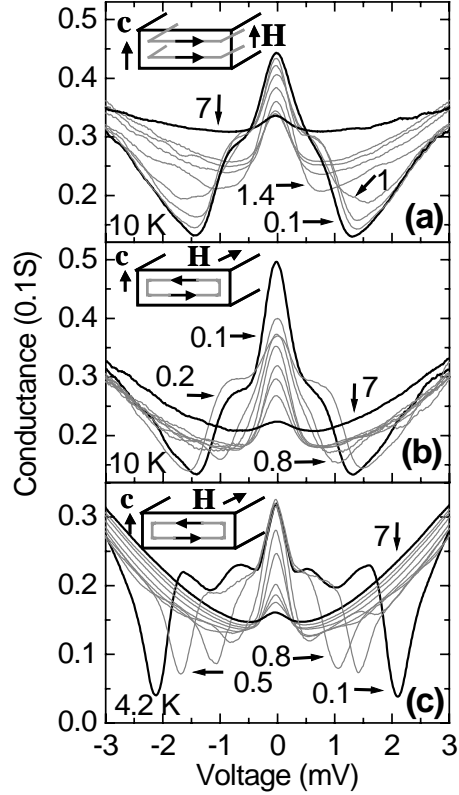

**FIGURE 3.** Variation of the G(V) spectra of GBJ #1 with magnetic field H ll *c*–axis (a) or H ll *ab* plane (b) at a temperature $T = 10$ K. For comparison in (c) some of the conductance spectra measured at 4.2 K for H ll *ab* plane are shown here again without shifting them vertically. In the insets the junction cross section is shown schematically together with the direction of screening current flow (horizontal arrows). The unspecified values of $\mu_0 H$ are: (a) 0.2, 0.8, 2, 3, 4 T; (b) 1, 2, 2.5, 3 T; (c) 1, 2, 3, 4, 5, 6 T. The numbers labeling the G(V) curves are field values in Tesla.

As in many other reports [5,8,13,21] for both temperatures (4.2 and 10 K) the ZBCP does not split in magnetic field as theoretically predicted [22]. A possible explanation for this behavior might be the considerable faceting of the grain-boundary which, together with impurity scattering, suppresses the field splitting of the ZBCP [2,3,13]. As predicted [22] the amplitude of the ZBCP decreases with increasing H (see Fig.2 and Fig.3). We found that at both temperatures this decrease is accompanied by a *nonlinear* reduction in the integrated density of states (IDOS) associated with the ZBCP (defined as $\int_{ZBCP} G(V)dV$), which is compensated by an increase in the IDOS at higher energy. As a result the IDOS from –3mV to 3mV (defined as $\int_{-3mV}^{3mV} G(V)dV$ ) remains almost unchanged. We illustrate this effect for the case of 10 K in Fig. 4, where full symbols are for $\int_{ZBCP} G(V)dV$, while empty symbols are for $\int_{-3mV}^{3mV} G(V)dV$. The effect is pronounced for fields up to 2 T when the IDOS associated with the ZBCP strongly decreases from its value at 0.1 T. For larger fields (in the range 2-7 T) the IDOS associated with the ZBCP saturates. This observation rules out the possibility that the observed ZBCP has a magnetic origin as in this case a *linear* displacement of states with increasing H and no saturation are expected [2,13]. Here by magnetic origin of the ZBCP we mean the Applebaum-Anderson mechanism [23] of inelastic tunneling via localized magnetic moments inside the barrier. As in the case of direct ZBCP splitting, the observed nonlinear energy displacement of states (also observed in hole doped HTS GBJs [12-13,18]) might be interpreted as a Doppler shift. The Doppler shift of the ABS energy [22] is $\mathbf{p}_S\mathbf{v}_F$, where $\mathbf{p}_S$, the superfluid momentum arising from the Meissner effect, and $\mathbf{v}_F$, the quasiparticle velocity at the Fermi level, are lying parallel to the *ab* planes. So far the Doppler shift effect observed in the junctions formed on top of (110)-oriented films was highly anisotropic [21,24,25]. It was strongest for H ll $c$ (screening currents at the junction interfaces flow in the *ab* plane) and much reduced for H ll $ab$ (screening currents flow along the $c$-axis). A strong anisotropy in the screening currents leads to the anisotropy of the Doppler shift. In contrast, for our junction geometry, independent of the direction of H (H ll $c$ or H ll $ab$), the most significant component of the screening currents are in the *ab* plane (see insets in Figs.2a, 2b, 3). Consequently, for both field orientations a considerable Doppler shift has to be present. This is exactly what we observe although, interestingly, there are differences (see Fig.4). Indeed, at 1 T for H ll $ab$ the IDOS associated with the ZBCP reduces down to about 50 % from its maximum while for H ll $c$ it only reaches its maximum. Then, for H ll $ab$ the IDOS associated with the ZBCP saturates at a lower value of about 35% from the value at 0.1 T (68 % for H ll $c$). Finally, as H increases for H ll $c$ the ZBCP's "center of mass" is located at larger conductances than for H ll $ab$ (compare Figs.3a and 3b). Equivalently, one can say that for all fields $\int_{ZBCP} G(V)dV$ is larger for H ll $c$ than for H ll $ab$ (see Fig. 4). We believe these difference are due to different screening current flow within the junctions in the two cases. Indeed, when H ll $c$, they have the same direction at every location within the junction (insets of Figs.2a and 3a) while, when H ll *ab*, the currents have opposite direction at the top and at the bottom of the junction (insets of Figs.2b and 3b).

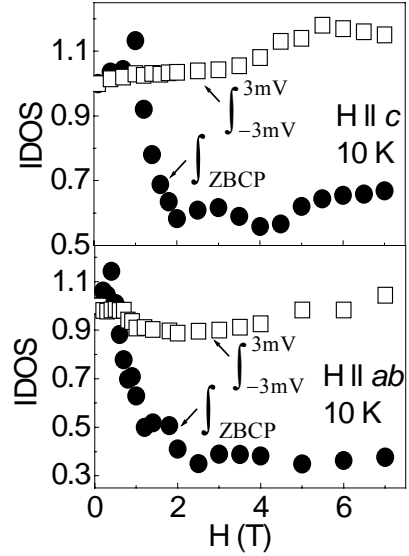

**FIGURE 4.** The integrated density of states (IDOS) versus H of the spectra shown in Fig.3. The IDOS is normalized to its value at 0.1 T. Full symbols are for IDOS associated with the ZBCP while empty symbols are for IDOS in the range from –3mV to 3 mV. The upper graph is for H ll $c$ axis, and the lower graph is for H ll $ab$ plane.

We have measured the temperature dependence of the ZBCP at fixed magnetic fields of 0.01, 0.1, 0.2, 0.3, 0.5, and 1 T with H ll *c*. Typical sets of such measurements (for 0.2 T and 1 T) are shown in Fig.5. The ZBCP as well as the gap structure (Fig. 5a) are suppressed with increasing temperature. The subgap background conductance gradually increases with increasing T consistent with a magnetic field induced reduction of the superconducting energy gap [26]. For all H measured, as T is increased from 4.2 K the amplitude of the ZBCP (defined as the difference between G(V=0) and the absolute minimum of the G(V)) first slightly increases and then monotonically decreases until it vanishes at $T_c$ (see inset of Fig.5a). On the other hand, if we decreased T from above the critical temperature $T_c$, for all values of applied field H, the ZBCP appears simultaneously with the onset of superconductivity (see inset of Fig.5a), i.e., at the critical temperature of about 29 K. If the ZBCP would have a magnetic origin [23] its amplitude would be a function of H and, as T is decreased, it should appear at different temperatures for different fields. This is not the case although H changes by 2 orders of magnitude from 0.01 T to 1 T. This strongly suggests that the ZBCP has not a magnetic origin [23] but is due to formation of ABS. The temperature dependence of the G(V) spectra is similar to other reports of ABS induced ZBCPs [21,27] and clearly shows that the superconducting state is being probed.

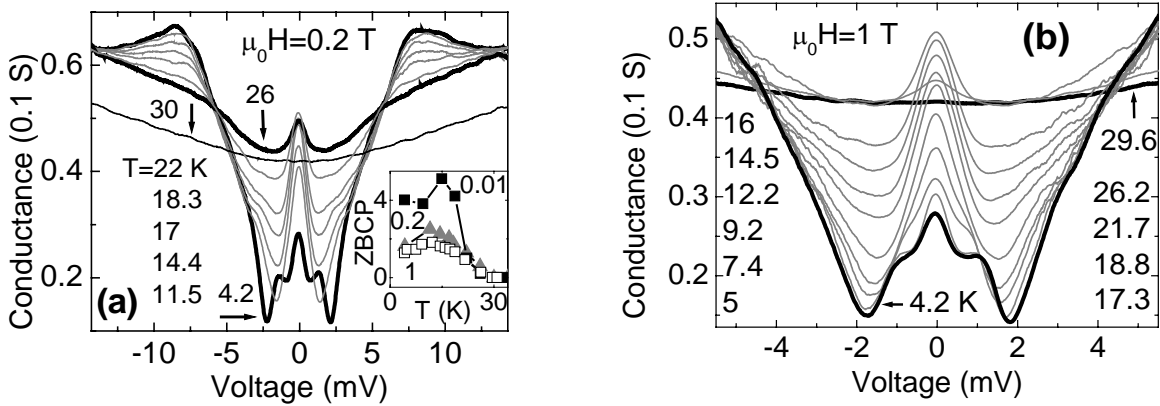

**FIGURE 5.** Variation of the G(V) spectra of GBJ #1 with temperature measured at a field of (a) 0.2 T and (b) 1 T applied parallel to the *c*-axis. The inset in (a) shows the ZBCP amplitude (in units of 0.01S) as a function of temperature for 3 different values of the applied field applied parallel to the *c*-axis: 0.01, 0.2, and 1 T.

Since we measured superconductor-insulator-superconductor (SIS) junctions it is possible that a Josephson supercurrent and not the formation of ABS might be responsible for the observed ZBCP. From the start it should be pointed out that it is very unlikely that at magnetic fields in the tesla range, where the ZBCP is very pronounced, there is any influence of the Josephson effect still present. Indeed, at fields in the mT range or higher there was no trace of a Josephson supercurrent left on the IVC. To be more convincing we investigated how the ZBCP presumably due to the Josephson tunneling changes with temperature and magnetic field. The junctions we measured are well described within the RCSJ model (see Fig.1). As we reduce $I_c$ by applying a magnetic field H or increasing temperature T, thermal fluctuations of energy $k_BT$ ($k_B$ is the Boltzmann constant) become important with respect to the Josephson coupling energy $E_J = I_c\Phi_0/2\pi$. Therefore, one should apply the RCSJ model in the presence of thermal fluctuations [28]. On the basis of this model that describes the behavior of the Josephson current at finite temperatures we calculated how the ZBCP, presumably due to the Josephson effect, changes with temperature and field. Figure 6 shows calculated Josephson conductance spectra for several values of the noise parameter $\gamma = E_J/k_BT = I_c\Phi_0/2\pi k_BT$ (note that frequently the noise parameter is defined as the inverse of the notation used here) and also (see the inset) how the corresponding

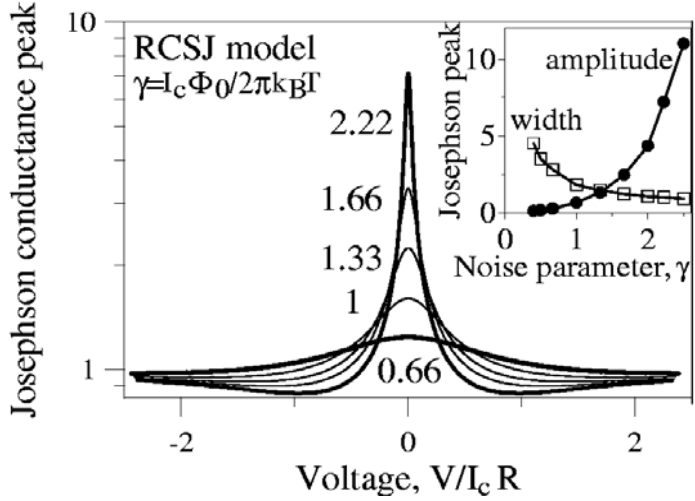

**FIGURE 6.** Calculated Josephson conductance peak as a function of the bias voltage for 5 different values of the noise parameter γ. The inset shows the amplitude and width of the Josephson conductance peak as a function the noise parameter γ.

ZBCP (amplitude and width) changes with γ. Within this model the amplitude of the Josephson conductance peak strongly increases with γ and its width monotonically decreases with γ (see Fig.6). In the experiments one reduces γ by increasing T or H (since both T and H suppress $I_c$). From Figs.2a, 2b and 3 it is clear that the width of the ZBCP first slightly increases, then decreases with increasing field. Then as T raises from 4.2 K up to 29 K (see Fig.5) the width of the ZBCP remains practically *unchanged* (initially it decreases and then increases back). As can be clearly seen by comparing Fig.3b with 3c by increasing T from 4.2 K to 10 K (we estimate that by doing that γ decreases by a factor of 3) the peak width gets smaller for small fields or remain practically unchanged for larger fields. All these observations are incompatible with the Josephson effect being the origin of the observed ZBCP. In addition, the zero-bias conductance (ZBC), i.e., the conductance at V = 0, increases with T (see Fig.5) which is exactly opposite of what is expected for a Josephson supercurrent induced ZBCP (see Fig.6; also [28]). Josephson supercurrents as the cause of the observed ZBCP can therefore be ruled out. On the other hand, a temperature-independent width of the ZBCP has been observed also in NIS junctions [21,27]. Such a behavior is consistent with the formation of ABS and it has been explained in terms of rough interfaces and umklapp surface scattering [29].

To get a qualitative theoretical understanding of the measured conductance spectra we have calculated the tunneling conductance of an SIS junction in the absence of an applied field using quasiclassical techniques as described in [22,30-34]. It allows us to calculate the local density of states at the two sides of the GBJ and the low transmission SIS normalized conductance

$$\frac{G(V)}{G_n} = \frac{1}{G_n}\frac{dI}{dV} = \frac{d}{dV}\int_{-\infty}^{\infty} d\omega N^l(\omega)N^r(\omega+V)[f(\omega)-f(\omega+V)] \quad (1).$$

Here $f(\omega)=1/(1+\exp(\omega\hbar/k_BT))$ is the Fermi distribution function. $N^l(\omega)$ and $N^r(\omega)$ are the (normalized) local densities of states in the superconducting state on the left and right hand side of the grain boundary. They are obtained using the Riccati method as described in Ref. [32, 33]. The finite barrier transmission is taken into account using Zaitsev's boundary conditions in the form of Eqs. (32) and (33) in Ref. [34]. Here we have used a finite barrier transmission coefficient of 0.1. In order to model elastic and inelastic scattering processes in the superconducting electrodes on both sides of the junction we have used a parabolic energy $\varepsilon$ dependent quasiparticle damping rate Γ of the form

$$\frac{\Gamma(\varepsilon)}{\Delta} = \frac{\Gamma_0}{\Delta} + \frac{1}{4}\left(\frac{\varepsilon}{\Delta}\right)^2 \tag{2}$$

in the calculation of $N^l(\omega)$ and $N^r(\omega)$, as shown in the inset of Fig. 7 for $\Gamma_0=0.125\Delta$. Thus, our calculation takes into account some GBJ features like a) tunneling in a SIS junction with a finite small barrier transmission; b) a $30^0$ misorientation angle at the GB; c) a pure *d*-wave symmetry for the superconducting order parameter; d) elastic and inelastic scattering at the interfaces. The results are shown in Fig.7 for $T/T_c=0.15$, and different values of the zero energy quasiparticle damping rate $\Gamma_0/\Delta$. The ZBCP is strongly suppressed by increasing the scattering rate (it vanishes when $\Gamma_0 = \Delta$), while for $\Gamma_0=0.225\Delta$ the peak height reproduces the experimental value.

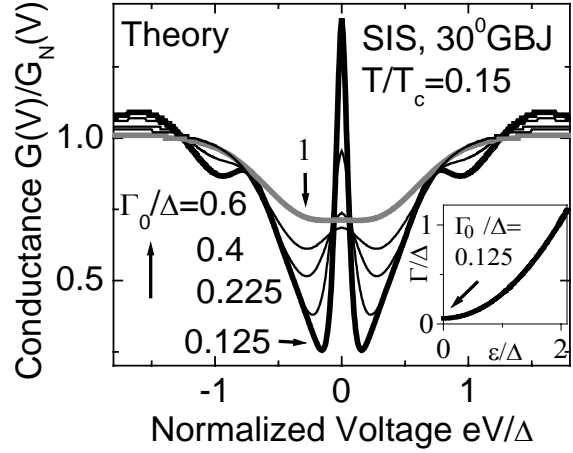

**FIGURE 7.** Calculated normalized $G(V)/G_N(V)$ spectra for 5 different values of the zero energy quasiparticle damping rate $\Gamma_0/\Delta$ and a barrier transmission coefficient of 0.1. The inset shows the parabolic energy $\varepsilon$ dependent quasiparticle damping rate used to calculated $G(V)/G_N(V)$ for $\Gamma_0/\Delta$ 0.125.

Why the ZBCP has not been previously observed in experiments involving electron doped GBJs [13-16]? As it is well known, electron-doped HTS GBJs have much smaller Josephson critical current densities $j_c$ as compared to hole doped HTS GBJs. To give an example, electron doped LCCO or NCCO $24^0$ GBJs have almost 6 orders of magnitude smaller $j_c$ than hole doped $YBa_2Cu_3O_7$ $24^0$ GBJs. That suggests that the electron-doped GBJs have a thicker junction barrier ($j_c$ exponentially decreases with the barrier thickness [14]). A thicker barrier means an enhanced scattering rate which is known to strongly reduce the ZBCP (see Fig.7 for SIS junctions; also [35], [36] for NIS junctions). It means, in principle, it should be more difficult to observe ABS induced ZBCP in electron-doped GBJs as compared to hole-doped GBJs. The electron doped $30^0$ GBJs we measured have $j_c$ between (6.3-11.5) A/cm$^2$ (cf. Table 1) comparable to reported values for electron doped $24^0$ GBJs, although $j_c$ exponentially decreases with the GB misorientation angle $\theta$: $j_c(\Theta) = j_c(\Theta=0)\exp(-\Theta/\Theta_i)$ [18,19]. Moreover, except for our group (this work; see also [9]), to our knowledge there have been no reports on electron doped $30^0$ GBJs having a measurable Josephson critical current density $j_c$. This proves a high quality of the GBJs we used (i.e., a high quality of both the STO substrate and the bicrystal line), translating into a thinner junction barrier and/or less disorder at the SIS interfaces. This means in our case there should be a significant reduction of the scattering rates (also due to a thinner junction barrier) and surface roughness which are known to strongly reduce the ZBCP (see Fig.7 for SIS junctions; also [35], [36] for NIS junctions).

In summary our results on magnetic field and temperature dependencies of the observed ZBCP in the tunneling spectra of LCCO GBJs strongly suggest that the origin of the observed peak is the formation of zero-energy Andreev bound states. This supports a predominantly *d*-wave symmetry of the order parameter in the nearly optimal doped LCCO cuprate. Taking into account our previous phase sensitive test based on the tunneling of Cooped pairs [9] the present work shows that both methods provide results that are consistent with each other. Consequently, our measurements solve the controversy between these two different types of phase sensitive tests previously performed with electron doped GBJs, namely, ABS-induced ZBCPs [13-16], and Josephson tunneling [1,9]. In the light of previous unsuccessful attempts to observe the ZBCP in electron doped HTS GBJs, the *observation* of the ZBCP rather than its *absence* should be regarded as a powerful tool to look into the symmetry of the order parameter in unconventional superconductors.

We thank M. Naito for his crucial support concerning film preparation and T. Nachtrab for his assistance on magnetic measurements. This work was supported by the ESF PiShift program and the Landesforschungsschwerpunktsprogramm Baden-Württemberg.